\documentstyle[preprint,prb,eqsecnum,aps%
,psfig,pstricks,multido,pst-coil]{revtex}              % new command
\begin{document}
\draft \flushbottom
%@twocolumnfalse\endcsname
\title{Multiple energy x--ray holography:\\
the polarization effect}
\author{V. Bortolani}
\address{INFM and Dipartimento di Fisica, Universit\`a di Modena,
via Campi 213/A, I-41100 Modena, Italy.}
\author{V. Celli}
\address{Physics Department, University of Virginia,
Charlottesville, Virginia 22904--4714.}
\author{A. M. Marvin}
\address{Dipartimento di Fisica Teorica, Universit\`a di Trieste,
Miramare--Grignano, I-34014 Trieste, Italy.}
\maketitle
\date{\today}
\begin{abstract}
We present the theory for Multiple Energy X--ray Holography 
(MEXH), using a multipole expansion for the scattered field. We 
find that light polarization plays a crucial role in the 
reconstruction of the image, and we suggest how to use it in order 
to eliminate aberration effects. The method we propose is 
alternative to the SWIFT method ({\em scattered-wave--included 
Fourier transform}), but has the advantage that no theoretical 
calculations are required to redefine the hologram. 
\end{abstract}
%
%]
%\narrowtext
\section{Introduction.}
The idea of holography dates back to Gabor's original 
works~\cite{1} and is essentially based on the interference of 
optical paths in crystal diffraction. The problem has been 
summarized very clearly by Sz\"oke~\cite{2}. Consider as a simple 
example, an emitter $A$ at the origin and a single neighbouring 
atom $N_1$ at a position ${\bf r}_1$. In the process known as {\em 
x--ray fluorescence holography} (XFH)~\cite{3,4}, the atom $A$ is 
excited and the emitted radiation either goes directly to the 
detector (photofilm) at a far distance $R$ (the so called {\em 
reference wave}), or is firstly scattered by $N$ ({\em object 
wave}). The photofilm registers in this way the interference of 
the two optical paths (hologram), and gets in Fourier transform
the desired information on the scatterer position ${\bf r}_1$.
Equivalently, as proposed originally by Gabor~\cite{1}, the last
step can be accomplished by illuminating the photographic film
with an incoming spherical wave~\cite{2,5} (the {\em decoding 
wave}), but numerical solutions that use a digital detector and
suppress the decoding wave are nowadays preferable.\par
\begin{figure}[ht]
\psset{unit=.3cm}
\begin{center}
\begin{pspicture}(-16,-4)(5,16)
%\psgrid
\newgray{mygray}{.9}
\psline[linewidth=.6pt](0,0)(-7,2)
\rput{40}(-7,2){%
\psline[linewidth=.6pt](0,0)(0,12) }
\rput{40}(0,0){%
\psline[linewidth=.6pt](0,0)(0,18) }
\rput{-50}(-11.57,13.79){%
\psCoil[coilaspect=0,coilheight=1.3,coilwidth=2,linewidth=1.pt%
,linestyle=dashed,dash=8pt 5pt,linecolor=gray]{90}{1710}
}
\rput{-50}(-14.7,11.19){%
\psCoil[coilaspect=0,coilheight=1.3,coilwidth=2,linewidth=1.pt%
,linestyle=solid,linecolor=gray]{90}{1550}
}
\rput{-15.94}(-3.5,1){%
\psCoil[coilaspect=0,coilheight=1.3,coilwidth=2,linewidth=1.pt%
,linestyle=solid,linecolor=gray]{-500}{270}
}
\pscircle[linestyle=none,fillstyle=solid,fillcolor=mygray](0,0){1}
\rput[bl](.8,.8){$A$}
\psline[linewidth=.4pt]{<->}(0,4)(0,0)(4,0)
\psline[linewidth=.4pt]{->}(0,0)(-2,-3.46)
\pscircle[linestyle=none,fillstyle=solid,fillcolor=gray](-7,2){1}
\rput[bl](-6.2,2.8){$N_1$}
\rput{-50}(-11.4,10.5){%
\psline[linewidth=1pt]{->}(0,0)(3,0)
\rput[l]{*0}(3.3,0){${\bf\hat k}$ }
}
\rput[bl](-16,13){\parbox[c]{2cm}{SR \\ source} }
\end{pspicture}
\end{center}
\caption{The picture of MEXH process.
The interference pattern is due to the different optical paths of
the direct beam (dashed) and the one scattered by the neighbouring
atom $N_1$ (solid line). \label{fig1} }
\end{figure}
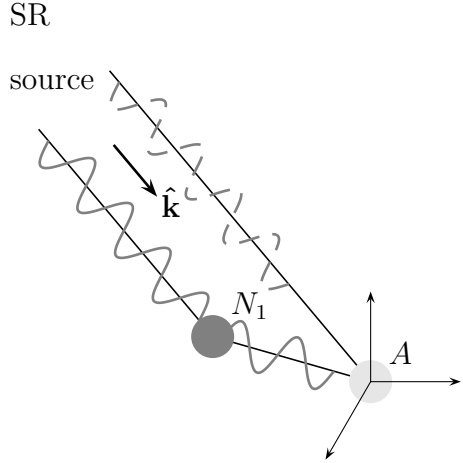

 
A step forward in this direction has been given recently by the group of
Fadley and Materlik~\cite{6}, in using  the emitter $A$ as a detector. 
In this experiment (MEXH), the reference wave is furnished by a
synchrotron radiation (SR) source, while the fluorescing atom $A$ 
senses an electric field which is the sum of the direct wave and
the scattered front (see Fig.~\ref{fig1}) due to its neighbouring
atoms ({\em object wave}).\par 
Formally, MEXH can be visualized as a time reversal situation of XFH, 
in the sense that the detector is substituted by the source, 
which obviously does {\em not} modify the hologram. In spite of this, 
it presents many hidden advantages.\par 
First of all, the hologram $\chi ({\bf k})$ is
defined at different energies. In MEXH and in contrast to XFH, the
photon energy is always above the fluorescence threshold of the
emitter, thus allowing a better resolution of the image as $\sim
k^{-1}$. Secondly, the SR photon energy in MEXH can be freely
varying, and this allows one to suppress unpleasant {\em
twin--image} effects, once the integration is not limited to a
sphere, but is done over a a 3--D  volume in ${\bf k}$
space~\cite{7}. A comparison between the two methods has been
given recently by Len {\em et al.}~\cite{8} and we refer to them
for further details.\par 
In this report we develop the full vector theory for MEXH. 
The general theory is presented in section~II. In
Sec.~III we make a multipole expansion of the previous formulae,
and discuss their limit of applicability. Here is where aberration
comes out, and we show in Sec.~IV how to avoid it, playing with
the polarization of the reference beam. Finally in Sec.~V we
confirm our previous findings presenting some numerical results on
a real crystal structure. Particular attention is devoted to the
disturbance induced by the scatterers at the large distance from
the emitter.
\section{The basic equations.} \label{s2}
Let us fix our attention on the neighbouring atoms first. The
external SR monochromatic field is represented by the {\em normalized} 
vector potential
\begin{equation}
{\bf A}({\bf r},\,t)=\mbox{\boldmath $\epsilon $}\,e^{i{\bf k\cdot 
r}} e^{-i\omega t}+\quad c.c. \label{21}
\end{equation}
where $k=\omega /c$ and $\mbox{\boldmath $\epsilon $}$ is the
polarization. This acts on each atom surrounding $A$ as a
disturbance $H_I\sim -e{\bf A\cdot p}$, causing a transition on
the electronic states. The density and current involved in this
process are
\begin{mathletters}
\begin{eqnarray}
\rho &\sim & -e\,\psi^*_n\psi_m +(m\leftrightarrow n)\label{22a}\\
{\bf j}&\sim & {-e\over m}\times\left[{-i\hbar\over 2}
\left(\psi^*_n\,{\bf\nabla}\psi_m -
\psi_m\,{\bf\nabla}\psi^*_n\right)\right. \nonumber \\
&&\left.\quad +(m\leftrightarrow n)+{e\over c}|\psi_n|^2\right]
,\label{22b}
\end{eqnarray}
\end{mathletters}
where $\psi_n$ is the initial core electron level, $\psi_m$ is the
final state, and the last term in (\ref{22b}) is the diamagnetic
contribution. Each transition is then weighted by appropriate
coefficients, which can be found by applying standard perturbation
theory.~\cite{9} Matching the density and the current density
operators between states, the result is of the same form as in
Eq.~(\ref{21}), i.e.
\begin{mathletters}
\begin{equation}
\rho ({\bf r},\,t)=\rho ({\bf r};\,\omega )e^{-i\omega t}+\quad
c.c.\label{23a}
\end{equation}
and
\begin{equation}
{\bf j}({\bf r},\,t)={\bf j}({\bf r};\,\omega )e^{-i\omega
t}+\quad c.c. \label{23b}
\end{equation}
\end{mathletters}
To find the two quantities on the r.h.s., we must pay attention
that: 1) the atoms are centered  at ${\bf r}_s$ and {\em not} 
at the origin; 2) the wavefunctions $\psi $  depend on the
relative position with respect to the nucleus
\begin{equation}
{\bf x}={\bf r}-{\bf r}_s \label{24}
\end{equation}
and are otherwise independent of $s$. This means that the induced
density (or current) for each atom, factorizes as a prefactor
$\exp (i{\bf k\cdot r}_s)$ times a term which is function of ${\bf
x}$ in Eq.~(\ref{24}) and  is {\em formally} independent of the atomic
position. The total induced density (or current) is the sum over
atoms. With this in mind the result is
\begin{equation}
\rho ({\bf r};\,\omega )=\sum_s e^{i{\bf k\cdot r}_s}\,\rho ({\bf
x}) \quad ;\quad {\bf j}({\bf r};\,\omega )=\sum_se^{i{\bf k\cdot
r}_s}\,{\bf j}({\bf x}) \label{25}
\end{equation}
where
\begin{eqnarray}
\rho ({\bf x})=-{e^2\over mc}\sum^3_{\beta =1}&&\sum_m\left[
-{{\psi^{(0)}_n}^*({\bf x})\psi^{(0)}_m({\bf x}) \langle m\vert
e^{i{\bf k\cdot x}^\prime}p_\beta\vert n\rangle \over \hbar
(\omega_{mn}-\omega -i\delta )}\right. \nonumber\\
&&{}\label{26}\\ &&\left. -{\langle n\vert e^{i{\bf k\cdot
x}^\prime}p_\beta\vert m\rangle {\psi^{(0)}_m}^*({\bf
x})\psi^{(0)}_n({\bf x}) \over \hbar (\omega_{mn}+\omega -i\delta
)}\right]\,\epsilon_\beta ,\nonumber
\end{eqnarray}
and
\begin{eqnarray}
j_\alpha ({\bf x})=&&-{e^2\over mc}\sum^3_{\beta =1}
\Bigg\{\delta_{\alpha\beta} e^{i{\bf k\cdot x}}
\vert\psi^{(0)}_n({\bf x})\vert^2 -{1\over m}\sum_m\Bigg[\nonumber
\\ && {{\psi^{(0)}_n}^*({\bf x})
\stackrel{\leftrightarrow}{p_\alpha} \psi^{(0)}_m({\bf x}) \langle
m\vert e^{i{\bf k\cdot x}^\prime}p_\beta\vert n\rangle \over \hbar
(\omega_{mn}-\omega -i\delta )} \label{27}\\ && +{\langle n\vert
e^{i{\bf k\cdot x}^\prime}p_\beta\vert m\rangle
{\psi^{(0)}_m}^*({\bf x})\stackrel{\leftrightarrow}{p_\alpha}
\psi^{(0)}_n({\bf x}) \over \hbar (\omega_{mn}+\omega -i\delta )}
\Bigg]\Bigg\}\,\epsilon_\beta .\nonumber
\end{eqnarray}
In Eq.~(\ref{27}) the dyadic over the momentum operator, ${\bf
p}=-i\hbar{\bf\nabla}$, means that it operates symmetrically both
on the right and on the left, according to the square bracket in
Eq.~(\ref{22b}). In the first row in Eq.~(\ref{27}) one recognizes
the diamagnetic contribution which does not depend on the virtual
states $m$, while in the denominators we have defined
\[
\omega_{mn}=\left(E^{(0)}_m-E^{(0)}_n\right)/\hbar ,
\]
and $E^{(0)}_{m,n}$ are energies of the excited and core electron
level. It can be verified from (\ref{26}) and (\ref{27}) that
Eq.~(\ref{23a}) and Eq.~(\ref{23b}) satisfy the charge
conservation ${\bf\nabla\cdot j}+\dot{\rho}=0$. Once the charges
and currents are given, the electric field at the emitter is
calculated from the Maxwell equations. Outside the sources, this
can be found from the vector potential only as
\begin{equation}
{\bf E}={i\over k}{\bf\nabla\times B}\quad ;\quad {\bf
B}={\bf\nabla\times A}\label{28}
\end{equation}
where
\[
{\bf A}({\bf r};\,\omega )={1\over c}\int {e^{ik|{\bf r}-{\bf
r}^\prime |}\over |{\bf r}-{\bf r}^\prime |} {\bf j}({\bf
r}^\prime ;\,\omega )\,d^3{\bf r}^\prime .
\]
Changing the integration variable as in Eq.~(\ref{24}), then using
the definition in (\ref{25}), one gets
\begin{eqnarray}
&&{\bf E}_{\rm obj}({\bf r};\,\omega )={i\over\omega} \sum_s
e^{i{\bf k\cdot r}_s}\Bigg\{ {\bf\nabla\times} \label{29}\\
&&\qquad \int\left[\left( {\bf\nabla}\strut{{
\displaystyle{e^{ik|{\bf r}-{\bf r}_s-{\bf x}^\prime |}} \over
|{\bf r}-{\bf r}_s-{\bf x}^\prime | }}\right){\bf\times j}({\bf
x}^\prime )\right]\,d^3{\bf x}^\prime \Bigg\} . \nonumber
\end{eqnarray}
The last equation is the object wave due to the scatterers. A
similar result has been obtained recently by Fonda~\cite{10} by
computing the field through the vector and scalar potentials, but
erroneously the latter was neglected from the start. Fonda's
result, in the present notation, is ${\bf E}_{\rm obj}=ik{\bf A}$,
which is equivalent to dropping one term of the double vector
product in Eq.~(\ref{29}). Further on, we show in Sec.~IV how this
correction becomes crucial for the polarization dependence of the
reconstructed image, which is the main point of this paper.\par
Adding from (\ref{21}) the direct contribution of the reference
wave, i.e.,
\begin{equation}
{\bf E}_{\rm ref}({\bf r};\,\omega )=ik\mbox{\boldmath $\epsilon
$}\, e^{i{\bf k\cdot r}},\label{210}
\end{equation}
we get the total field
\begin{equation}
{\bf E}_{\rm Tot}={\bf E}_{\rm ref}+{\bf E}_{\rm obj}\label{211}
\end{equation}
that acts as a perturbation on the fluorescing atom $A$.
\section{The multipole expansion.}
The states of $A$ involved in the transition are again core
electron levels. This means that one needs Eq.~(\ref{211}) for
${\bf r}={\bf x}$ and $|{\bf x}|\alt d$, with $d$ of the order of
the Bohr radius. In the same way, the integral in Eq.~(\ref{29}) is
confined to $|{\bf x}^\prime |\alt d$, i.e., in the volume where
the current in (\ref{22b}) is appreciably different from zero.
Since $r_s$ is of the order of the lattice parameter, it follows that
\[
r_s\gg x,\,x^\prime .
\]
This allows us to use for the propagator in Eq.~(\ref{29}) the so
called plane wave approximation (PWA)~\cite{10}
\begin{equation}
{e^{ik|{\bf r}-{\bf r}_s-{\bf x}^\prime |}\over |{\bf r}-{\bf
r}_s-{\bf x}^\prime |}\approx {e^{ik|{\bf r}-{\bf r}_s|}\over
|{\bf r}-{\bf r}_s|}\, e^{-i{\bf k}_s{\bf\cdot x}^\prime
}\label{31}
\end{equation}
with ${\bf k}_s=-k{\bf\hat r}_s$ the scattered momentum. One gets
\begin{eqnarray}
&&{\bf E}_{\rm obj}({\bf x};\,\omega )=\sum_s k^2 e^{i{\bf
k}_s{\bf\cdot x}}\Bigg\{ [{\bf M}_s-({\bf M}_s{\bf\cdot\hat
r}_s){\bf\hat r}_s]\nonumber \\ && +{1\over ikr_s}[3({\bf
M}_s{\bf\cdot\hat r}_s){\bf\hat r}_s-{\bf M}_s] \left(1-{1\over
ikr_s}\right)\Bigg\}\,{e^{ikr_s}\over r_s}\, e^{i{\bf k\cdot
r}_s}\label{32}
\end{eqnarray}
with
\begin{equation}
{\bf M}_s={i\over\omega}\int e^{-i{\bf k}_s{\bf\cdot x}^\prime}
{\bf j}({\bf x}^\prime )\,d^3{\bf x}^\prime .\label{33}
\end{equation}
Eq.~(\ref{32}) has been written for completeness. For $kd\ll 1$,
$\exp (-i{\bf k}_s{\bf\cdot x}^\prime )\approx 1$ in
Eq.~(\ref{33}) and ${\bf M}_s\approx {\bf p}$,~\cite{11} with
\begin{equation}
{\bf p}=\int {\bf x}^\prime \rho ({\bf x}^\prime )\,d^3{\bf
x}^\prime \label{34}
\end{equation}
the electric dipole moment, and $\rho $ the density in Eq.~(26). In this
case the object field in (\ref{32}) is constant over the atom. In
the two extreme limits $kr_s\gg 1$ and $kr_s\ll 1$ one discovers
in it the familiar result (see Ref.~[\onlinecite{11}]) of the
radiative and the near static field in the dipole
approximation.\par
For x--rays the limit $kr_s\gg 1$ is appropriate, but $k\sim d^{-1}$
is required for good resolution. This means
that under the sum in (\ref{32}) only the first term is
contributes, while in Eq.~(\ref{33}) one is led to attempt an
expansion as
\[
e^{-i{\bf k}_s{\bf\cdot x}^\prime }\approx 1-i{\bf k}_s{\bf\cdot
x}^\prime +\cdots
\]
The result is
\begin{eqnarray}
&&{\bf E}_{\rm obj}({\bf x};\,\omega )=\sum_s k^2e^{i{\bf
k}_s{\bf\cdot x}} \Bigg\{\left[-{\bf\hat r}_s{\bf\times}({\bf\hat
r}_s{\bf\times p})\right] \nonumber \\ && +\left[{\bf\hat
r}_s{\bf\times m}\right]-{i\over 3!} \left[-{\bf\hat
r}_s{\bf\times}({\bf\hat r}_s{\bf\times Q}_s)\right]\Bigg\}
{e^{ikr_s}\over r_s}e^{i{\bf k\cdot r}_s}\label{35}
\end{eqnarray}
where ${\bf m}$ is the magnetic dipole moment,
\begin{mathletters}
\begin{equation}
\left({\bf Q}_s\right)_\alpha =\sum^3_{\beta =1} Q_{\alpha\beta}
\left({\bf k}\right)_\beta ,\label{36a}
\end{equation}
and $Q_{\alpha\beta}$ is the electric quadrupole moment tensor
\begin{equation}
Q_{\alpha\beta}=\int \left(3x^\prime_\alpha x^\prime_\beta
-{r^\prime}^{\,2}\delta_{\alpha\beta}\right)\,\rho ({\bf x}^\prime
)\, d^3{\bf x}^\prime .\label{36b}
\end{equation}
\end{mathletters}
To make the expansion consistent, we shall suppose in the
following that the last two terms in Eq.~(\ref{35}) are small.
Equally well we shall suppose that the same happens for the
fluorescing atom $A$, such that the limit ${\bf x}=0$ in Eq.~(35)
is appropriate. The yield is simply $\sim\vert{\bf E}({\bf
x}=0)\vert^2$ and function of the direction ${\bf\hat k}$ of the
laser beam only. The hologram follows as
\begin{mathletters}
\begin{eqnarray}
\chi ({\bf k})=&&\left(\vert{\bf E}_{\rm Tot}\vert^2 -\vert{\bf
E}_{\rm ref}\vert^2\right)/\vert{\bf E}_{\rm ref}\vert^2
\label{37a} \\ &&\approx {1\over k^2}\left( {\bf E}^*_{\rm
ref}{\bf\cdot E}_{\rm obj}\,+\quad c.c.\right)\label{37b}
\end{eqnarray}
\end{mathletters}
and in the last line the {\em self--hologram} $\sim {\bf E}^2_{\rm
obj}$ has been neglected. Using (\ref{210}) and (\ref{34}) the
result for Eq.~(\ref{37b}) is
\begin{equation}
\chi ({\bf k})=-ik\sum_sf_s{e^{ikr_s}\over r_s}e^{i{\bf k\cdot
r}_s} \,+\quad c.c.\label{38}
\end{equation}
and where $f_s$ the scattering amplitude
\begin{eqnarray}
f_s =&&\left(\mbox{\boldmath $\epsilon $}^*{\bf\times\hat
r}_s\right) {\bf\cdot}\left({\bf p\times\hat r}_s\right)
+\mbox{\boldmath $\epsilon $}^*{\bf\times r}_s{\bf\cdot
m}\nonumber \\ &&\qquad\quad -{i\over 3!} \left(\mbox{\boldmath
$\epsilon $}^*{\bf\times r}_s\right){\bf\cdot} \left({\bf
Q}_s{\bf\times\hat r}_s\right).\label{39}
\end{eqnarray}
Equation above shows that $f_s$ is complex and is polarization
dependent. The quadrupole correction, causes a shift~\cite{8},
while the dipole terms give rise to a distortion of the image
({\em aberration effects}). In practical cases (see
Ref.~[\onlinecite{8}] for details) the shift is $\sim 1/10$ of the
resolution and thus negligible. On the contrary the distortion is
effective and is polarization dependent.
\section{Angular anisotropies.}
For a constant $f_s$ in Eq.~(\ref{38}), the Fourier transform
\begin{equation}
U({\bf r};\,k)={1\over 4\pi}\int \chi ({\bf k}) e^{-i{\bf k\cdot
r}}\,d{\bf\hat k},\label{41}
\end{equation}
with $d{\bf\hat k}$ denoting the solid angle $d\Omega_{{\bf k}}$,
solves completely the problem of the atomic position. In this
case~\cite{2}
\begin{eqnarray}
U({\bf r};\,k)=&&-ik\sum_s f_s{e^{ikr_s}\over r_s}\, {\sin (k|{\bf
r}-{\bf r}_s|)\over k|{\bf r}-{\bf r}_s|}\nonumber \\ && +ik\sum_s
f^*_s{e^{-ikr_s}\over r_s}\, {\sin (k|{\bf r}+{\bf r}_s|)\over
k|{\bf r}+{\bf r}_s|}\label{42}
\end{eqnarray}
and shows spherical illuminated spots with resolution $\sim
k^{-1}$ centered at the atoms and their {\em twins}. The twins do
not represent a problem in MEXH. In fact, as suggested by
Barton~\cite{7}, they can be mostly eliminated by integrating
Eq.~(\ref{41}) over the energy as
\[
U({\bf r})=\int U({\bf r};\,k)e^{-ikr}\,k^2 dk
\]
and we shall not discuss it. We shall concentrate on how to render
$f_s$ constant.\par 
One way is to use the SWIFT method proposed by Saldin {\em et al.}~\cite{12}. 
In it, the Fourier transform Eq.~(\ref{41}) is done {\em  not} on the 
{\em bare} hologram $\chi ({\bf k})$ as one gets from the data, 
but on the redefined quantity
\begin{equation}
\chi_{\rm SWIFT}({\bf k};\,{\bf r})={\chi ({\bf k})\over f_{{\bf
r}}} \label{43}
\end{equation}
where
\[
f_r=f_s({\bf\hat r}_s\rightarrow {\bf\hat r})
\]
and $f_s$ is defined in Eq.~(\ref{39}). Using (\ref{38}) one notes
that as ${\bf r}$ in (\ref{41}) approaches ${\bf r}_s$,
${f_s/ f_r}\approx 1$, which guarantees that
both shifts and distortions are eliminated in Eq.~(\ref{41}).\par
The method applies for any polarization, but problems may arise
when the ratio $f_s/f_{{\bf r}}$ is $\sim 0/0 $. In addition the
dipole moments and the quadrupole term are assumed to be
known.\par 
Another way to render $f_s$ constant, would be instead
to use the bare experimental quantity $\chi ({\bf k})$, but to
play with the polarization.\par 
Neglect for the moment the corrections in (\ref{39}) and concentrate 
on the first term only. To make the life easier, 
here and in the following we shall suppose that ${\bf p}$ is in the 
direction of the field. From this we define
\begin{equation}
f_s\sim \vert \mbox{\boldmath $\epsilon $}{\bf\times\hat
r}_s\vert^2 =\sin^2\Psi, \label{44}
\end{equation}
and $f_s=1$, if
\begin{equation}
\mbox{\boldmath $\epsilon $}\rightarrow\mbox{\boldmath $\epsilon
$}_s ={{\bf\hat r}_s{\bf\times\hat k}\over |{\bf\hat
r}_s{\bf\times\hat k}|}. \label{45}
\end{equation}
The choice made in Eq.~(\ref{45}) implies the {\em a priori}
knowledge of the position ${\bf r}_s$, but this can be roughly
estimated through Eq.~(\ref{41}) and with an arbitrary
polarization. The method we propose is the following.\par 
Let $\mbox{\boldmath $\epsilon $}_1$, and 
$\mbox{\boldmath $\epsilon$}_2 
={\bf\hat k\times}\mbox{\boldmath $\epsilon $}_1$ be two
polarizations where the second is obtained by a rotation of 
$\pi /2$ over the direction ${\bf\hat k}$ of the reference beam.
Measure the yield for $\mbox{\boldmath $\epsilon $}_1$,
$\mbox{\boldmath $\epsilon $}_2$ and for
\[
\mbox{\boldmath $\epsilon $}_3={1\over\sqrt{2}}\left(
\mbox{\boldmath $\epsilon $}_1+\mbox{\boldmath $\epsilon
$}_2\right)
\]
then call $\chi_1$, $\chi_2$ and $\chi_3$ the respective
holograms. The third measurement collects interference effects and
constructs the hologram $\chi_{(\alpha ,\beta )}$ for any
polarization
\begin{mathletters}
\begin{equation}
\mbox{\boldmath $\epsilon $}=\alpha\mbox{\boldmath $\epsilon $}_1
+\beta\mbox{\boldmath $\epsilon $}_2\quad ;\quad |\alpha
|^2+|\beta |^2 =1 \label{46a}
\end{equation}
to be used in the experiment. For instance if both $\alpha $ and
$\beta $ are real one gets from (\ref{37a}) the simple result
\begin{equation}
\chi_{(\alpha ,\beta )}=(\alpha -\beta )
(\alpha\chi_1-\beta\chi_2)+2\alpha\beta\chi_3 .\label{46b}
\end{equation}
\end{mathletters}
Taking for $\alpha $ and $\beta $ the values
\begin{equation}
\alpha_s={{\bf\hat r}_s{\bf\cdot}\mbox{\boldmath $\epsilon
$}_2\over |{\bf\hat r}_s{\bf\times\hat k}|}\quad ,\quad
\beta_s=-{{\bf\hat r}_s{\bf\cdot}\mbox{\boldmath $\epsilon
$}_1\over |{\bf\hat r}_s{\bf\times\hat k}|},\label{47}
\end{equation}
$\mbox{\boldmath $\epsilon $}=\mbox{\boldmath $\epsilon $}_s$,
thus $f_s=1$ as wanted.\par 
We show now with an example how, without corrections, 
distortion is effectively present. Suppose
that ${\bf r}_s$ is along the $z$ axis and take
\[
\mbox{\boldmath $\epsilon $}_1={{\bf\hat z\times\hat k}\over
|{\bf\hat z\times\hat k}|}
\]
in the $(x,y)$ plane. Then $\vert\mbox{\boldmath $\epsilon
$}_1{\bf\times r}_s\vert^2 =1$, but
\[
\vert\mbox{\boldmath $\epsilon $}_2{\bf\times r}_s\vert^2
=\cos^2\vartheta
\]
where
\[
{\bf\hat k}=(\sin\vartheta\cos\varphi ,\,\sin\vartheta\sin\varphi
,\,\cos\vartheta ).
\]
Spherical spots are insured for $\mbox{\boldmath $\epsilon $}_1$,
but not for $\mbox{\boldmath $\epsilon $}_2$ as we now show.\par
Consider two paths: the first with ${\bf r}$ along $z$, and the
second with ${\bf r}-{\bf r}_s$ in the $(x,y)$ plane. For both
paths the function $U$ in Eq.~(\ref{41}) depends only on
\[
\zeta =k|{\bf r}-{\bf r}_s|.
\]
A simple calculation for the first path gives
\[
U_\parallel (\zeta
)={1\over\zeta}\left[\left(1-{2\over\zeta^2}\right) \sin\zeta
+{2\over\zeta}\cos\zeta\right]
\]
with a maximum $U_\parallel (0)=1/3$, and vanishing at
\[
\zeta\approx {\pi\over 1.5}.
\]
For the second path $U_\perp (0)=U_\parallel (0)$, but the shape
is wider and vanishes at
\[
\zeta\approx {\pi\over .757}.
\]
In this case then the illuminated spot is a highly anisotropic
ellipse. The two axes of the ellipse are in the ratio
$1.5/.757\approx 2$.
\section{Numerical results and conclusions.}
A numerical calculation has been carried out for a Fe bcc
structure, to see how the method works. With the emitter at the
origin, we limit our attention on the nearest neighbour spot ${\bf
r}^*=({1\over 2}{1\over 2}{1\over 2})a $, where $a=2.87$ \AA is the
lattice spacing for iron. The plane is the $(1\bar{1}0)$ one and
the directions $[110]$ and $[001]$ as indicated in the figures.\par 
Fig.~\ref{fig2a} is for polarization $\mbox{\boldmath
$\epsilon $}_1$, here both orthogonal to ${\bf\hat r}^*$ and
${\bf\hat k}$ as in Eq.~(\ref{45}). A spherical spot is clearly
shown, in spite of the possible disturbances induced by other
scatterers, and neglected in the analysis of the previous section.
Fig.~\ref{fig2b} refers to $\mbox{\boldmath $\epsilon $}_2$
polarization, gotten by rotating $\mbox{\boldmath $\epsilon $}_1$ by
$\pi /2$ around ${\bf\hat k}$ as indicated previously. Here the
intensity is lowered and the anisotropy is present indeed. Both
figures are for reference beam energy $E=6.24$~KeV. The
beam energy cannot be arbitrary, but must be chosen to avoid out of
phase overlap between an atom and its twin, a well known
problem in single-energy holography\cite{8}. In fact the sum in
Eq.~(\ref{38}) is invariant for ${\bf r}_s\rightarrow -{\bf r}_s$,
thus one gets (apart from a $2k^2$ factor) the dimensionless
quantity
\begin{equation}
\chi ({\bf k})=\sum_sf_s{\sin (kr_s)\over kr_s}\cos ({\bf k\cdot
r}_s), \label{51}
\end{equation}
with $f_s\le 1$ for both polarizations given in Eq.~(\ref{44}). 
The prefactor $P=\sin (kr_s)/ kr_s $ selects the possible energies for the 
image reconstruction as $\sin (kr^*)\approx +1$. Reversing  the sign 
in defining the $U$ function in Eq.~(\ref{41}), the values 
$\sin (kr^*)\approx -1$ are again possible, but we exclude them since are 
{\em minima} in the $|{\bf E}_{\rm Tot}|^2$ intensity in Eq.~(\ref{37a}). 
In the present case the most suitable values for the image reconstruction
for $r^*$-scatterer are $E_2=6.24$~KeV and $E_3=11.23$~KeV (the
first value $E_1=1.25$~KeV being too small).\par
Figures~\ref{fig3a} and \ref{fig3b} refer to this higher energy
$E_3$. Again one notices that the $r^*$-atom is visualized only
for $\mbox{\boldmath $\epsilon $}_1$ polarization
(Fig.~\ref{fig3a}), while this spot vanishes using the
polarization $\mbox{\boldmath $\epsilon $}_2$
(Fig.~\ref{fig3b}).\par 
To get closer to the experimental conditions, 
the $\chi $ values in Eq.~(\ref{51}) have been
calculated on a $5^\circ\times 5^\circ $-grid over the whole solid
angle, then averaged over an experimental window. A 
{\em spline interpolation} on these grid--data creates the averaged
hologram to be used under the integral in Eq.~(\ref{41}). In all
cases we find that the angular averaging is irrelevant as far as
$U({\bf r})$ is concerned. The primary role is played by the
integration over the modulus, i.e. over the energy. Here we use a
1--D gaussian convolution $g\sim \exp (-k^2/{\Delta k}^{\,2})$ on
the grid data. The effect of using the gaussian is twofold, and
well discussed in Ref.~[\onlinecite{3}]. On the one hand it is
nothing but the energy spread of the primary beam. On the other,
it acts as a low--pass filter to wash out all the distant
scatterers, and at the same time avoids divergences in the
reconstructed pattern. The first point is closely related to the
dimension of the cluster, or {\em mutatis mutandis}, to the number
of atoms of a {\em bulk} crystal we include under the sum in
Eq.~(\ref{38}). We use a sphere of radius $R=a\times N$ and find
that, for $\Delta k\approx\pi/(aM)$ and $N>M$, 
the result is independent of the sphere's radius.\par
There is however a divergence for large $N$, which  has more to
do with our definition of $U$
than with the physics itself. It can be avoided when
the integration is {\em not} performed overthe {\em whole} solid
angle as in Eq.~(\ref{41}), but only on a part of it. This is
what is usually done by the people working in atomic
holography~\cite{13,14}. Nevertheless we shall continue to use
Eq.~(\ref{41}), and show how to get rid of this problem.\par 
Its origin can be figured out very simply by going to the scalar case,
thus taking $f_s=1$ constant. From Eq.~(\ref{42}) get the
intensity at the origin
\[
U(0)\rightarrow \sum_s {\sin^2(kr_s)\over k^2r^2_s}\sim R
\]
which diverges for $R\rightarrow\infty $. The same trend is of
course present for $|{\bf r}|\approx 0$, and this means a very
faint intensity in the whole pattern, except at the emitter. However,
this divergence can be removed in the calculation by subtracting
a $C$--constant, just as the {\em reference wave} is subtracted to
get at the experimental $\chi $. The $C$ value we use is fixed by
taking constant $U(0)$ as the sphere radius is increased from
$R=a$ to $R=Na$. This is a very good way to check how the result
stabilizes with increasing number of scatterers, even if other
choices are possible~\cite{13}.\par 
We find much faster convergence at lower energy, as one would expect. 
The fourier transform (FT) $U({\bf r})$ is smoother and the illuminated 
spots are larger.\par 
Actually for $E=E_2$ we do {\em not} even need to
perform a gaussian convolution, but only a {\em C--redefinition}
of the type discussed is required. Nevertheless an energy
spread is used in all the figures we present. We fix $\sigma
={\Delta k/ k}=6.411\times 10^{-2}$, close to the Tegze and
Feigel\cite{3} value, this being the {\em default} value for
$R\agt 3a$, and energy $E_3$. We have checked, as just noted, that
Figs.~\ref{fig2a} and \ref{fig2b} remain in practice the same
using a larger $\sigma $ or setting  $\Delta k =0$ (monocromatic
beam). On the contrary, in Fig.~\ref{fig3a} and Fig.~\ref{fig3b} a
gaussian convolution is essential.\par 
The numerical calculation is performed with a {\em Fortran} program. 
Once the averaged holograms $\chi_1$, $\chi_2$ and $\chi_3$ 
are calculated on the $5^\circ\times 5^\circ $ grid, 
the FT requires 2' and 30" of computer time for each reconstruction 
at $E=11.23$~KeV. For Figs.~\ref{fig2a}, \ref{fig2b} the requirement 
reduces to $\sim 1'$. 
For the graphical part we use {\em Mathematica}. To get
better contrast in the figures, we follow the standard procedure
used in atomic holography~\cite{14}, namely plotting $|U|^2$
rather then $|U|$. Before doing this $U$ has been redefined such
that  $0\le U\le 1$ in the whole $0\le x\le\sqrt{2}a$, $0\le y\le
a$ $(1\bar{1},0)$ plane, or in the relevant part of it.\par 
In conclusion, we can assert that the reconstruction of the image can
be gotten with a reasonable effort from the experimental hologram.
The energy spread of the nearly monochromatic reference beam,
prevents to see distant scatterers, but justifies to neglect the
{\em self-hologram} contribution. An additional convolution of the
data with a gaussian lowers the definition of the image, but makes
the {\em self-hologram} even smaller. The polarization, together
with the energy, plays an essential role for the observability of
the atom. The polarization `~{\em constraint}~' remains effective
even if, as in MEXH, the integration over energy is performed.\par
\acknowledgments{One of us (A. M. M.) would like to thank L. Fonda
for stimulating discussions on this subject. We thank J.S. Frasier for a
critical reading of the manuscript and help with editing it.}
\newcounter{subfigure}
\setcounter{subfigure}{1}
\renewcommand{\thefigure}{\arabic{figure}\alph{subfigure}}
\begin{figure}[ht]
\psfig{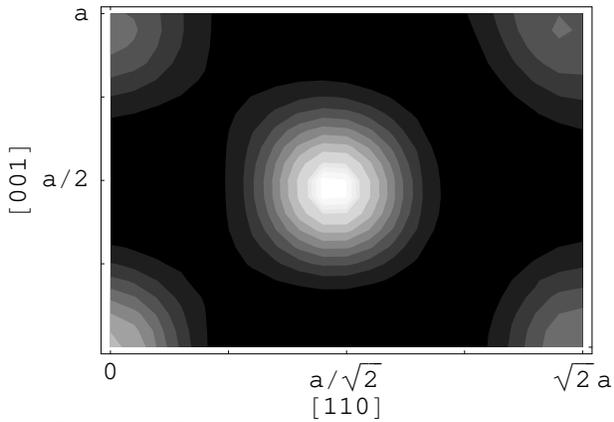}
\caption{Theoretically generated atomic images for a $Fe $ bcc
crystal on the $(1\bar{1}0)$ plane. The X-ray energy is
$E=6.24$~KeV and  the polarization $\mbox{\boldmath $\epsilon
$}_1$  is defined in the text. Other parameters are $R=3a$ (258
scatterers), $C=.273$, and ${\Delta k/ k}=.064$
for the gaussian convolution.\label{fig2a} }
\end{figure}
\stepcounter{subfigure}
\addtocounter{figure}{-1}
\begin{figure}[ht]
\psfig{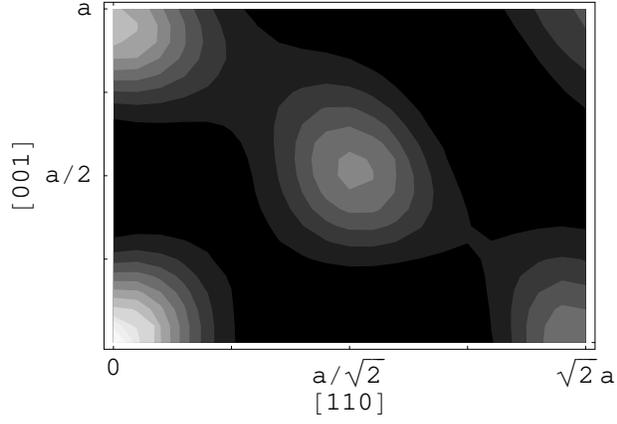}
\caption{As in Fig.~\ref{fig2a}, but for the $\mbox{\boldmath
$\epsilon $}_2$ polarization.\label{fig2b} }
\end{figure}
\setcounter{subfigure}{1}
\begin{figure}[ht]
\psfig{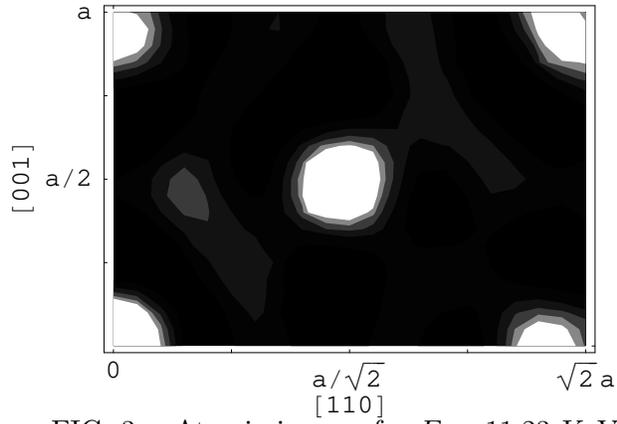}
\caption{Atomic images for $E=11.23$~KeV and polarization $\mbox{\boldmath
$\epsilon $}_1$. Here we have used $R=4a$ (536 scatterers),
$C=.132$, and the same ${\Delta k/k}$.\label{fig3a} }
\end{figure}
\stepcounter{subfigure}
\addtocounter{figure}{-1}
\begin{figure}[ht]
\psfig{figure=fig3b.epsi}
\caption{As in Fig.~\ref{fig3a}, but for $\mbox{\boldmath
$\epsilon $}_2$ polarization.\label{fig3b} }
\end{figure}
\end{document}